\documentstyle[11pt,epsf]{article}
\begin{document}
\title{On the semiclassical treatment of anharmonic quantum oscillators via
coherent states -- The Toda chain revisited}
\author{John Schliemann and Franz G. Mertens\\
{\it Physikalisches Institut, Universit{\"a}t Bayreuth, D-95440 Bayreuth,
Germany}}
\date{March 1999}
\maketitle
\begin{abstract} 
We use coherent states as a time--dependent variational {\it ansatz}
for a semiclassical treatment of the dynamics of anharmonic 
quantum oscillators. In this approach the square variance of the Hamiltonian 
within coherent states is of particular interest. 
This quantity turns out to have a natural interpretation
with respect to time--dependent solutions of the semiclassical equations
of motion. Moreover, our approach allows for an estimate of the 
decoherence time of a classical object due to quantum fluctuations.
We illustrate our findings at the example of the Toda chain.
\end{abstract}


\section{Introduction}
\label{1}

Coherent states are an important notion in quantum physics, in particular
with respect to semiclassical approximations; for general references
see \cite{Per:86,KlSk:85,ZFG:90}.\\
The coherent states of the harmonic oscillator have been introduced by
Schr\"odinger \cite{Sch:26} and have been reexamined by Glauber \cite{Gla:63} 
in circumstances of quantum optics. For spin systems, spin--coherent states,
i.~e. the coherent states of SU(2), have been introduced by Radcliffe
\cite{Rad:71}. These two types of coherent states provide an immediate
connection to the classical limit of generic quantum systems and are the most
important examples of coherent states in physics.\\
The connection to the classical limit is obtained by using coherent
states as a time--dependent variational {\it ansatz} to investigate the 
dynamics of a quantum system. Recently, this approach has been  reconsidered
by the present authors with respect to interacting spin systems given by
a general Heisenberg model \cite{ScMe:98}. The central result in that work is 
the evaluation of the square variance of the Hamiltonian within coherent
states. This quantity turns out to have a natural interpretation with
respect to time--dependent spin structures and allows also for an estimate
of the validity of the variational approach. In the present work we extend
these results to the case of oscillator systems.\\
In classical nonlinear lattices and as well in classical spin
systems certain nonlinear excitations like solitary waves are of particular
interest. However, it is an open question whether such dynamic and spatially
localized excitations can also exist in the corresponding quantum systems.
The results of this work provide an estimate for the lifetime of
such objects. We demonstrate this for the example of the Toda chain.\\ 
The outline of this paper is as follows: In section \ref{2} we summarize
the essential properties of the coherent states of the harmonic oscillator,
and in section \ref{3} we introduce the time--dependent variational method
in quantum mechanics. This method in used in the next section to treat a 
generic anharmonic oscillator. In particular, the square variance of the
Hamiltonian is evaluated. This quantity shows very analogous properties 
to those obtained in \cite{ScMe:98} for the case of quantum spin systems.
These findings can be extended to the case of several anharmonically coupled
degrees of freedom; as an example we examine the quantum Toda chain
in sections \ref{5} and \ref{6}.


\section{Coherent states of the harmonic oscillator}
\label{2}

The Hamiltonian of the quantum harmonic oscillator is given in standard
notation by
\begin{equation}
{\cal H}_{h}=\frac{p^{2}}{2m}+\frac{m\omega^{2}}{2}q^{2}
=\hbar\omega\left(a^{+}a+\frac{1}{2}\right)
\label{hH}
\end{equation}
with
\begin{equation}
a=\frac{1}{\sqrt{2}}\left(\sqrt{\frac{m\omega}{\hbar}}q
+\frac{i}{\sqrt{\hbar m\omega}}p\right)\quad,\quad 
a^{+}=\left(a\right)^{+}
\label{aa}
\end{equation}
and the well--known commutation relations
\begin{equation}
\left[p,q\right]=\frac{\hbar}{i}
\quad\Leftrightarrow\quad\left[a,a^{+}\right]=1\,.
\label{cr}
\end{equation}
The quantities $\sqrt{\hbar/m\omega}$ and $\sqrt{\hbar m\omega}$ arising in
the operators (\ref{aa}) are the characteristic length and momentum,
respectively. The system has an equidistant spectrum. Eigenstates are
naturally labelled by $n\in\{0,1,2,\dots\}$,
\begin{equation}
{\cal H}_{h}|n\rangle=\hbar\omega\left(n+\frac{1}{2}\right)|n\rangle\,.
\end{equation} 
Coherent states of the harmonic oscillator are eigenstates of the lowering
operator $a$ with complex eigenvalues $\alpha$,
\begin{equation}
a|\alpha\rangle=\alpha|\alpha\rangle\,.
\label{defcs}
\end{equation}
They can be expressed as
\begin{equation}
|\alpha\rangle=\exp\left(\alpha a^{+}-\alpha^{\ast}a\right)|0\rangle
=\exp\left(-\frac{1}{2}|\alpha|^{2}\right)
\sum_{n=0}^{\infty}\frac{\alpha^{n}}{\sqrt{n!}}|n\rangle\,.
\label{cs1}
\end{equation}
The parameter $\alpha$ is naturally decomposed into its real and imaginary
part as
\begin{equation}
\alpha=\frac{1}{\sqrt{2}}\left(\sqrt{\frac{m\omega}{\hbar}}\xi
+\frac{i}{\sqrt{\hbar m\omega}}\pi\right)\,.
\end{equation}
Denoting an expectation value within a coherent state (\ref{cs1}) by
$\langle\cdot\rangle$ it holds
\begin{equation}
\langle q\rangle=\xi\quad,\quad\langle p\rangle=\pi\,.
\label{exval1}
\end{equation}
Coherent states maintain their shape in the time evolution of the harmonic
oscillator,
\begin{equation}
e^{-\frac{i}{\hbar}{\cal H}_{h}t}|\alpha\rangle
=e^{-\frac{i}{2}\omega t}|\alpha e^{-i\omega t}\rangle\,,
\end{equation}
and the time dependence of the expectation values (\ref{exval1}) follows
exactly the classical motion of the harmonic oscillator. This fact justifies
the term `coherent states'. A further important property of these
objects is their completeness,
\begin{equation}
\frac{1}{\pi}\int d^{2}\alpha|\alpha\rangle\langle\alpha|=1\,,
\end{equation} 
but it should be mentioned that an arbitrary linear combination ot coherent
states has not the property (\ref{defcs}). Thus, the coherent states do not 
form a subspace of the Hilbert space but rather a submanifold.


\section{The time--dependent variational method}
\label{3}

The Schr\"odinger equation of quantum mechanics can be derived by extremizing
the action functional
\begin{equation}
{\cal S}=\int_{t_{i}}^{t_{f}}dt
\langle\psi|i\hbar\frac{d}{dt}-{\cal H}|\psi\rangle
\label{var1}
\end{equation}
with respect to the quantum state $|\psi(t)\rangle$ (or $\langle\psi(t)|$)
which is kept fixed at the times $t_{i}$ and $t_{f}$ \cite{KrSa:81}.
An approximate
approach to the dynamics of a quantum system can be performed by restricting
the states in (\ref{var1}) to a certain submanifold of the Hilbert space.
In the context of semiclassical approximations coherent states are a natural
choice. E.~g. for a single particle moving in a potential the appropriate
objects are coherent oscillator states as described in the foregoing section.
Thus, our restricted action functional reads in this case
\begin{equation}
\tilde{\cal S}=\int_{t_{i}}^{t_{f}}dt
\langle\alpha|i\hbar\frac{d}{dt}-{\cal H}|\alpha\rangle
=\int_{t_{i}}^{t_{f}}dt\left(\pi\partial_{t}\xi
-\langle{\cal H}\rangle\right)\,,
\label{var2}
\end{equation} 
where we have left out a total time derivative in the last integrand.
The coherent state $|\alpha\rangle$ is employed here as a time--dependent
variational {\it ansatz}, i.~e. its time dependence is assumed to be given by
time--dependent parameters $\pi(t)$, $\xi(t)$.
This restricted variational principle can be recognized as the stationary
phase condition for the quantum mechanical transition amplitude between
fixed  states $|\alpha(t_{i})\rangle$ and $|\alpha(t_{f})\rangle$
when expressed as a path integral over coherent states \cite{Kla:78},
\begin{equation}
{\cal U}(t_{i},t_{f})=\int{\cal D}\alpha
\exp\left(\frac{i}{\hbar}\int_{t_{i}}^{t_{f}}dt
\langle\alpha|i\hbar\frac{d}{dt}-{\cal H}|\alpha\rangle\right)\,.
\end{equation}
The variational equations of motion obtained from (\ref{var2}) are
\begin{equation}
\partial_{t}\xi=\frac{\partial\langle{\cal H}\rangle}{\partial\pi}
\quad,\quad
\partial_{t}\pi=-\frac{\partial\langle{\cal H}\rangle}{\partial\xi}\,,
\label{eom1}
\end{equation}
which have the same form as the classical Hamilton equations.\\
The time--dependent variational {\it ansatz} of coherent states becomes
exact if the potential in the Hamiltonian is harmonic.
Therefore, our approximate description of the quantum dynamics should be
valid for not too large anharmonicities. This will be examined further
in the next section. 


\section{The anharmonic oscillator}
\label{4}

Let us consider a generic anharmonic quantum oscillator
\begin{equation}
{\cal H}=\frac{p^{2}}{2m}+\frac{m\omega^{2}}{2}q^{2}+\frac{a}{3}q^{3}
+\frac{b}{4}q^{4}\,.
\label{aos}
\end{equation}
With coherent states as a time--dependent variational ansatz we
find for the expectation value of the energy
\begin{eqnarray}
\langle{\cal H}\rangle & = & \frac{1}{2m}\left(\pi^{2}+
\frac{1}{2}\hbar m\omega\right)
+\frac{m\omega^{2}}{2}\left(\xi^{2}+
\frac{1}{2}\frac{\hbar}{m\omega}\right)\nonumber\\
& & +\frac{a}{3}\left(\xi^{3}+\frac{3}{2}\xi\frac{\hbar}{m\omega}\right)
+\frac{b}{4}\left(\xi^{4}+3\xi^{2}\frac{\hbar}{m\omega}
+\frac{3}{4}\left(\frac{\hbar}{m\omega}\right)^{2}\right)\,,
\label{energy1}
\end{eqnarray}
which is of course constant in time. The variational equations of motion
(\ref{eom1}) read 
\begin{eqnarray}
\partial_{t}\xi & = & \frac{\pi}{m}\,,
\label{eom2}\\
\partial_{t}\pi & = & -m\omega^{2}\xi 
-a\left(\xi^{2}+\frac{1}{2}\frac{\hbar}{m\omega}\right)
-b\left(\xi^{3}+\frac{3}{2}\frac{\hbar}{m\omega}\xi\right)\,.
\label{eom3}
\end{eqnarray}
It is worthwhile to note that the same equations can be obtained from
the Heisenberg equations of motion for the operators $q$ and $p$,
\begin{equation}
\partial_{t}q=\frac{i}{\hbar}\left[{\cal H},q\right]\quad,\quad
\partial_{t}p=\frac{i}{\hbar}\left[{\cal H},p\right]\,,
\end{equation}
when the expectation values of both sides of the equations are taken within 
the state $|\alpha\rangle$ and the same assumption about its time evolution 
is made as above. This approach has been used by Krivoshlykov {\it et al.}
\cite{KMS:82}.\\
The equations (\ref{energy1})--(\ref{eom3}) reduce to the classical ones in 
the limit $\hbar\to 0$. Therefore, the coherent states reproduce the
classical limit.\\
Next let us examine the square variance of the energy, i.~e.
$\langle{\cal H}^{2}\rangle-\langle{\cal H}\rangle^{2}$. This quantity is
non--zero only in the quantum case and, as well as $\langle{\cal H}\rangle$,
strictly an invariant of the system, whatever the exact quantum mechanical
time evolution of the coherent state is. The square variance can be written
in the form
\begin{equation}
\langle{\cal H}^{2}\rangle-\langle{\cal H}\rangle^{2}=\Omega_{1}+\Omega_{2}
\label{sv1}
\end{equation}
with
\begin{equation}
\Omega_{1}=\frac{1}{2}\left(
\left(\hbar m\omega\right)\left(\frac{\pi}{m}\right)^{2}
+\frac{\hbar}{m\omega}\left(m\omega^{2}\xi 
+a\left(\xi^{2}+\frac{1}{2}\frac{\hbar}{m\omega}\right)
+b\left(\xi^{3}+\frac{3}{2}\frac{\hbar}{m\omega}\xi\right)\right)^{2}\right)\,,
\end{equation}
\begin{equation}
\Omega_{2}=\frac{1}{2}\left(\frac{\hbar}{m\omega}\right)^{2}
\left(a\xi+\frac{3}{2}b\xi^{2}\right)^{2}
+\left(\frac{\hbar}{m\omega}\right)^{3}
\left(\frac{a^{2}}{12}+2b^{2}\xi^{2}+\frac{5}{4}ab\xi\right)
+\left(\frac{\hbar}{m\omega}\right)^{4}b^{2}\frac{3}{8}\,.
\end{equation}
The quantity $\Omega_{1}$ is of leading order $\hbar$, while $\Omega_{2}$
contains only higher orders. The squared expressions in $\Omega_{1}$ can
be recognized as the right hand sides of (\ref{eom2}), (\ref{eom3}).
Thus, we have
\begin{equation}
\Omega_{1}=\frac{1}{2}\left(
\left(\hbar m\omega\right)\left(\partial_{t}\xi\right)^{2}
+\frac{\hbar}{m\omega}\left(\partial_{t}\pi\right)^{2}\right)\,.
\label{Omega1}
\end{equation}
Within our variational approach, the first order in $\hbar$ of the square 
variance
of the Hamiltonian is purely due to the time dependence of the state vector.
On the other hand, for a quantum state which has a non--trivial time evolution
and is consequently not an eigenstate of the Hamiltonian, the energy must
definitely have a finite uncertainty. Following this observation, the
first order in (\ref{sv1}) is not to be considered as a artifact of
our variational {\it ansatz}, but as a physically relevant expression for
the uncertainty of the energy for a time dependent solution to the
variational equations of motion (\ref{eom2}), (\ref{eom3}). Therefore,
the variational approach with coherent states does not only reproduce
the classical limit, but is also meaningful for a semiclassical
description of the anharmonic oscillator.\\
The contributions of higher order summarized in $\Omega_{2}$ indicate
limitations of our variational {\it ansatz},
i.~e they are a measure of decoherence effects due to the quantum mechanical
time evolution. To clarify this, let us consider the temporal
autocorrelation function
\begin{equation}
\langle\alpha|e^{-\frac{i}{\hbar}{\cal H}t}|\alpha\rangle\,,
\label{sc}
\end{equation} 
i.~e. the projection of the time--evolved state onto the initial coherent
state. The modulus of this quantity depends on time for two different
reasons: Firstly the quantum state has a non--trivial semiclassical
time evolution described by the equations (\ref{eom2}), (\ref{eom3}).
In real space the coherent state is represented by a Gaussian. Within our
{\em semiclassical} description of the dynamics the wave function remains a 
Gaussian, but its center is moving. Therefore the overlap of the
initial state and the time--evolved state is reduced 
Secondly, defects of our variational approach, which lead to 
decoherence effects,
also diminish the scalar product (\ref{sc}). Such quantum fluctuations 
affect the shape of the wave function which will not remain strictly of the 
Gaussian form under the {\em exact} quantum mechanical time evolution in the 
anharmonic potential.
The latter effects become 
significant on a time scale given by the uncertainty relation, where the
relevant contribution to the uncertainty of the energy is given by
$\Omega_{2}$,
\begin{equation}
\sqrt{\Omega_{2}}\Delta t\geq\frac{\hbar}{2}\,.
\label{ur}
\end{equation}
Alternatively one may consider the following correlation amplitude
\begin{equation}
C(t):=\langle\alpha(t)|e^{-\frac{i}{\hbar}{\cal H}t}|\alpha\rangle
\label{coramp}
\end{equation}
with $\alpha(t)$ given by time--dependent functions $\xi(t)$ and $\pi(t)$
which are solutions of (\ref{eom2}), (\ref{eom3}) with the initial
condition $\alpha(0)=\alpha$. This quantity is the projection of the
coherent state evolved under the exact quantum mechanical time evolution
onto the state given by the semiclassical time evolution. If the
potential in the Hamiltonian is purely harmonic we have $|C(t)|=1$ and 
$\Omega_{2}$ vanishs. In this case our variational {\it ansatz} of coherent 
states is of course exact and no 
decoherence effects occur. This observation also supports our
interpretation of the different contributions to 
$\langle{\cal H}^{2}\rangle-\langle{\cal H}\rangle^{2}$.\\
Thus deviations of the
modulus of (\ref{coramp}) from unity measure decoherence effects due
to the exact quantum mechanical time evolution under the anharmonic 
Hamiltonian. These effects manifest themselves in the additional
contribution $\Omega_{2}$ to the square variance of the energy.
The leading order $\Omega_{1}$ can be interpreted purely as an effect of
the semiclassical time evolution which does not incorporate decoherence
effects since it assumes the state vector to remain within the submanifold
of coherent states throughout the time evolution.\\
If one inserts a generic time--dependent solution $\xi(t)$, $\pi(t)$  
of the semiclassical equations of motion (\ref{eom2}), (\ref{eom3}) in
$\Omega_{1}$ and $\Omega_{2}$ these quantities will not be constant in time
separately (although their sum $\Omega_{1}+\Omega_{2}$ stricly is constant
in the exact quantum mechanical time evolution).
However, as an approximation, one may use in (\ref{ur}) the value of
$\Omega_{2}$ given by the initial value of $\xi$. This is justified if the
semiclassical motion of the particle is not too fast, i.~e. the semiclassical
momentum $\pi$ is not too large. In particular, if the initial coherent state
is chosen to have $\pi(0)=0$ and a certain value of $\xi$, the particle
will move in the semiclassical description to smaller $\xi(t)$ because of the
attractive potential. In this case the $\Omega_{2}$ evaluated for
the initial value $\xi(0)$ is an upper bound for $\Omega_{2}$
evaluated for later times, since this quantity grows with increasing $\xi$.
Reversely speaking, quantum fluctuations summarized in the quantity 
$\Omega_{2}$ become larger if the $\xi$ approaches the turning point of the
semiclassical motion governed by the equations (\ref{eom2}), (\ref{eom3}).
This feature is well--known from the usual WKB--approximation and therefore
consistent with the interpretation of $\Omega_{2}$ given above. 
Moreover, in the following we will also examine other systems  which exhibit 
stationary semiclassical dynamics with $\Omega_{1}$ and $\Omega_{2}$ being 
constant in time separately.\\
Another example where the validity of our considerations can be checked
explicitely is the free particle with ${\cal H}=p^{2}/2m$. Let the 
particle be initially in a coherent state with the wave function
\begin{equation}
\langle q|\alpha\rangle=\left(\frac{m\omega}{\pi\hbar}\right)^{\frac{1}{4}}
\exp\left(-\frac{m\omega}{2\hbar}\left(q-\xi\right)^{2}
+\frac{i}{\hbar}\pi\left(q-\frac{\xi}{2}\right)\right)\,.
\end{equation}
The quantity $\omega$ is not a frequency here but a parameter which
determines the localization of the particle in real and momentum space
around the expectation values (\ref{exval1}). The square variance of the
Hamiltonian reads the same as in (\ref{sv1}) with
\begin{equation}
\Omega_{1}=\frac{1}{2}\hbar m\omega\left(\frac{\pi}{m}\right)^{2}\quad,\quad
\Omega_{2}=\frac{1}{8}\left(\hbar\omega\right)^{2}\,.
\end{equation}
 Since the expectation value of the momentum is constant for such a 
translationally invariant system, $\Omega_{1}$ and $\Omega_{2}$ are
conserved separately.
The time--evolved wave function can be obtained readily as
\begin{equation}
\langle q|e^{-\frac{i}{\hbar}{\cal H}t}|\alpha\rangle=
\left(\frac{m\omega/{\pi\hbar}}{1+(\omega t)^{2}}\right)^{\frac{1}{4}}
\exp\left(-\frac{m\omega}{2\hbar}
\frac{\left(q-\xi-\frac{\pi}{m}t\right)^{2}}{1+(\omega t)^{2}}\right)
e^{i\varphi(q,t)}
\end{equation}
with a real phase $\varphi(q,t)$. Thus, the width of the wave function
increases, i.~e. its spatially localized structure is smeared out, on a time
scale of $\Delta t=1/\omega$, which is consistent with the estimate
given by (\ref{ur}). This result also strongly supports the above 
interpretation of the quantities $\Omega_{1}$ and $\Omega_{2}$.\\
In the next section we will make further use of the estimate of the 
decoherence time $\Delta t$ provided by (\ref{ur}).\\
The findings described above are completely analogous to the results 
obtained recently on interacting spin systems with spin--coherent states
as a time--dependent variational {\it ansatz} \cite{ScMe:98}. The
particular case corresponding to the harmonic limit of an oscillator
is given here by a paramagnet, where all spins are independent of
each other and coupled only to a static magnetic field. In this case all
spins perform a Larmor precession around the field axis, and this motion is
described exactly by spin--coherent states. A further common aspect of the
harmonic oscillator and a spin in a magnetic field is the equidistance
of the spectra of both systems.


\section{Anharmonic lattices: The Toda chain}
\label{5}

It is an obvious idea to generalize the results of the foregoing section
to systems with many anharmonically coupled degrees of freedom.
Let us consider an Hamiltonian ${\cal H}={\cal T}+{\cal V}$ with
\begin{equation}
{\cal T}=\sum_{n=0}^{N-1}\frac{p_{n}^{2}}{2m}\quad,\quad
{\cal V}=\sum_{n=0}^{N-1}V(q_{n}-q_{n-1})\,,
\end{equation}
where $N$ is the number of degrees of freedom and periodic boundary 
conditions are imposed. The 2--particle potential $V(x)$
contains in general anharmonic terms. To give a semiclassical description
of the dynamics, one may proceed similarly as for the single anharmonic
oscillator, but for a general potential $V(x)$ such an approach
leads to quite complicated expressions, in particular for the square
variance of the energy. Fortunately, a special case
exists where the results can be given in a concise form. This case is
the Toda chain, which is well--known in the theory of nonlinear lattices
\cite{Tod:81},
\begin{equation}
V(x)=\frac{\eta}{\gamma^{2}}\left(e^{-\gamma x}+\gamma x -1\right)
=\frac{m\omega^{2}e^{-\gamma\lambda}}{\gamma^{2}}
\left(e^{-\gamma x}+\gamma x -1\right)\,.
\label{pot}
\end{equation}
The potential $V$ contains the two parameters $\eta$ and $\gamma$; for
further convenience we have rewritten $\eta$ in terms of the new parameters
$\omega$ and $\lambda$ to be determined below. In the limit $\gamma\to 0$
the system is just the harmonic chain having independent phonon
modes labelled by the wave number $k$ with the acoustic phonon 
dispersion $\omega(k)=2\omega\sin(|k|/2)$. The usual phonon operators read
\begin{equation}
b_{k}=\frac{1}{\sqrt{2N}}\sum_{n=0}^{N-1}
\left[\left(\sqrt{\frac{m\omega(k)}{\hbar}}q_{n}
+\frac{i}{\sqrt{\hbar m\omega(k)}}p_{n}\right)e^{-i kn}\right]
\quad,\quad b_{k}^{+}=(b_{k})^{+}\,.
\end{equation}
An appropriate variational {\it ansatz} is given by coherent phonon states,
\begin{equation}
|\beta\rangle=\bigotimes_{k\in {\rm 1.BZ}}|\beta_{k}\rangle\,,
\label{ansatz}
\end{equation}
where the coherent state of the mode $k$ fullfills 
$b_{k}|\beta_{k}\rangle=\beta_{k}|\beta_{k}\rangle$ and the tensor product
runs over the first Brillouin zone. Again we denote expectation values 
within (\ref{ansatz}) by $\langle\cdot\rangle$. The parameters $\beta_{k}$
are related to the local expectation values $\langle q_{n}\rangle=\xi_{n}$,
$\langle p_{n}\rangle=\pi_{n}$ by
\begin{equation}
\beta_{k}=\frac{1}{\sqrt{2N}}\sum_{n=0}^{N-1}
\left[\left(\sqrt{\frac{m\omega(k)}{\hbar}}\xi_{n}
+\frac{i}{\sqrt{\hbar m\omega(k)}}\pi_{n}\right)e^{-i kn}\right]\,.
\label{copar}
\end{equation}
Such an approach to the dynamics of the
quantum Toda chain has been performed by Dancz and Rice \cite{DaRi:77},
and by G\"ohmann and Mertens \cite{GoMe:92}. Here we add
instructive results on the square variance of the Hamiltonian.\\
The expectation value of the Hamiltonian reads
\begin{equation}
\langle{\cal H}\rangle=\sum_{n=0}^{N-1}\frac{\pi_{n}^{2}}{2m}
+\sum_{n=0}^{N-1}\frac{m\omega^{2}}{\gamma^{2}}
e^{-\gamma\left(\lambda-\frac{\gamma}{2}\Delta_{0}\right)}
\left(e^{-\gamma\left(\xi_{n}-\xi_{n-1}\right)}-1\right)\,,
\label{ev2}
\end{equation}
where $\Delta_{0}$ is a correlation in the phononic vacuum $|0\rangle$. 
More generally, one has
\begin{eqnarray}
\Delta_{p} & := & \langle 0|\left(q_{n+p}-q_{n+p-1}\right)
\left(q_{n}-q_{n-1}\right)|0\rangle\nonumber\\
 & = & \frac{\hbar}{m\omega}\frac{1}{2N}
\frac{-\sin\left(\frac{\pi}{N}\right)}
{\sin\left(\frac{2p+1}{2N}\pi\right)\sin\left(\frac{2p-1}{2N}\pi\right)}
\,{{{N\to\infty}\atop{\longrightarrow}}\atop{}}\,
\frac{-1}{(4p^{2}-1)}\frac{2}{\pi}\frac{\hbar}{m\omega}\,,
\end{eqnarray}
where the following relations hold:
\begin{equation}
\sum_{p=0}^{N-1}\Delta_{p}=0\quad,\quad
\sum_{p=0}^{N-1}\left(\Delta_{p}\right)^{2}=\frac{1}{2}
\left(\frac{\hbar}{m\omega}\right)^{2}\,.
\label{sums}
\end{equation}
The expectation value (\ref{ev2}) has the same form as the classical Toda 
Hamiltonian up to a renormalization of the parameter $\lambda$. The 
equations of motion are obtained analogously as in (\ref{eom1}) and have
therefore also the same functional form as the classical ones. 
It was shown in \cite{GoMe:92} that this is a peculiarity of the Toda 
potential.\\
From the equations of motion one obtains
\begin{eqnarray}
\sum_{k}\left[\hbar^{2}\left(\partial_{t}\beta_{k}\right)
\left(\partial_{t}\beta_{k}^{\ast}\right)\right] & = &
\sum_{n,n'}\Biggl[
\frac{\pi_{n}}{m}\left(m^{2}\omega^{2}\Delta_{n-n'}\right)
\frac{\pi_{n'}}{m}\nonumber\\
 & & \quad+\left(\frac{m\omega^{2}}{\gamma}
e^{-\gamma\left(\lambda-\frac{\gamma}{2}\Delta_{0}\right)
-\gamma\left(\xi_{n}-\xi_{n-1}\right)}\right)
\left(\Delta_{n-n'}\right)\nonumber\\
 & & \qquad\cdot
\left(\frac{m\omega^{2}}{\gamma}
e^{-\gamma\left(\lambda-\frac{\gamma}{2}\Delta_{0}\right)
-\gamma\left(\xi_{n'}-\xi_{n'-1}\right)}\right)
\Biggr]\,.
\label{eom4}
\end{eqnarray}
Note that the left hand side of (\ref{eom4}) is of leading order $\hbar$,
since the parameters $\beta_{k}$ contain a factor $1/\sqrt{\hbar}$ 
(cf.~(\ref{copar})).\\
The square variance of the Hamiltonian reads
\begin{equation}
\langle{\cal H}^{2}\rangle-\langle{\cal H}\rangle^{2}
=R_{1}+R_{2}+R_{3}
\end{equation}
with
\begin{eqnarray}
R_{1} & = & \langle{\cal T}^{2}\rangle-\langle{\cal T}\rangle^{2}\nonumber\\
 & = & \sum_{n,n'}\Bigl[
\frac{\pi_{n}}{m}\left(m^{2}\omega^{2}\Delta_{n-n'}\right)
\frac{\pi_{n'}}{m}\Bigr]+\frac{N}{4}(\hbar\omega)^{2}\,,
\label{R1}\\
R_{2} & = & \langle{\cal V}^{2}\rangle-\langle{\cal V}\rangle^{2}\nonumber\\
 & = & \sum_{n,n'}\Biggl[\left(\frac{m\omega^{2}}{\gamma^{2}}
e^{-\gamma\left(\lambda-\frac{\gamma}{2}\Delta_{0}\right)
-\gamma\left(\xi_{n}-\xi_{n-1}\right)}\right)
\left(e^{\gamma^{2}\Delta_{n-n'}}-1\right)\nonumber\\
 & & \qquad\cdot
\left(\frac{m\omega^{2}}{\gamma^{2}}
e^{-\gamma\left(\lambda-\frac{\gamma}{2}\Delta_{0}\right)
-\gamma\left(\xi_{n'}-\xi_{n'-1}\right)}\right)
\Biggr]\,,
\label{R2}\\
R_{3} & = & \langle{\cal TV}+{\cal VT}\rangle
-2\langle{\cal T}\rangle\langle{\cal V}\rangle\nonumber\\
 & = & -(\hbar\omega)^{2}\frac{1}{2}\sum_{n}\Bigl[
e^{-\gamma\left(\lambda-\frac{\gamma}{2}\Delta_{0}\right)
-\gamma\left(\xi_{n}-\xi_{n-1}\right)}\Bigr]\,.
\label{R3}
\end{eqnarray}
These expressions can be derived by similar methods as described in 
\cite{GoMe:92}.
The technical advantage of the Toda potential lies in the fact that the 
contribution $R_{2}$ has a comparatively simple form and can be obtained
via the Baker--Campbell--Hausdorff--identity.\\
Expanding the factor $(\exp(\gamma^{2}\Delta_{n-n'})-1)$ in (\ref{R2}) and
using the equations (\ref{eom4}), (\ref{sums}) one can rewrite these formulae 
as
\begin{equation}
\langle{\cal H}^{2}\rangle-\langle{\cal H}\rangle^{2}
=\sum_{\mu=1}^{\infty}\Omega_{\mu}
\label{Osum}
\end{equation}
with
\begin{eqnarray}
\Omega_{1} & = & \sum_{k}\left[\hbar^{2}\left(\partial_{t}\beta_{k}\right)
\left(\partial_{t}\beta_{k}^{\ast}\right)\right]\,,
\label{O1}\\
\Omega_{2} & = & \frac{1}{2}\sum_{n,n'}
\Biggl[\left(\frac{m\omega^{2}}{\gamma^{2}}
\left(e^{-\gamma\left(\lambda-\frac{\gamma}{2}\Delta_{0}\right)
-\gamma\left(\xi_{n}-\xi_{n-1}\right)}-1\right)\right)
\left(\gamma^{2}\Delta_{n-n'}\right)^{2}\nonumber\\
 & & \qquad\cdot
\left(\frac{m\omega^{2}}{\gamma^{2}}
\left(e^{-\gamma\left(\lambda-\frac{\gamma}{2}\Delta_{0}\right)
-\gamma\left(\xi_{n'}-\xi_{n'-1}\right)}-1\right)\right)
\label{O2}
\Biggr]\,,
\end{eqnarray}
and for $\mu>2$
\begin{eqnarray}
\Omega_{\mu} & = & \sum_{n,n'}\Biggl[\left(\frac{m\omega^{2}}{\gamma^{2}}
e^{-\gamma\left(\lambda-\frac{\gamma}{2}\Delta_{0}\right)
-\gamma\left(\xi_{n}-\xi_{n-1}\right)}\right)
\frac{1}{\mu!}\left(\gamma^{2}\Delta_{n-n'}\right)^{\mu}\nonumber\\
 & & \qquad\cdot
\left(\frac{m\omega^{2}}{\gamma^{2}}
e^{-\gamma\left(\lambda-\frac{\gamma}{2}\Delta_{0}\right)
-\gamma\left(\xi_{n'}-\xi_{n'-1}\right)}\right)
\Biggr]\,.
\label{O3}
\end{eqnarray}
Each term $\Omega_{\mu}$ is of leading order $\hbar^{\mu}$
because $\Delta_{n-n'}\propto\hbar$. As seen from
(\ref{O1}) the lowest order in $\hbar$ in the square variance of the 
Hamiltonian is purely given by the time dependence of the semiclassical
variables. Therefore, the same conclusions apply as in the foregoing section.
Note also that again in the harmonic limit $\gamma\to 0$ all $\Omega_{\mu}$ for
$\mu>1$ vanish and the variational {\it ansatz} is exact.\\
We have demonstrated the result given in the equations (\ref{Osum}),
(\ref{O1}) for the Toda chain as an example, mostly to reduce technical
difficulties. In fact, from the experience with an analogous semiclassical
treatment of quite general Heisenberg spin models in arbitrary
spatial dimension \cite{ScMe:98}, these findings are 
expected to hold for more general lattice models.


\section{Decoherence effects to semiclassical solitary waves in the Toda
chain}
\label{6}

In the last decades an immense literature has emerged
on solitons in solid state physics. In those publications, the solid is usually
modelled (at least effectively)
as a classical system, while in fact it carries generally
quantum degrees of freedom. We will see below how our approach can be used to 
make contact between the classical and the quantum mechanical description.
In particular, the validity of theories based on classical solitary
excitations can be estimated.\\
The one--dimensional Toda lattice is an integrable system in the classical
\cite{Tod:81,cltoda} as well as in the quantum mechanical case \cite{qmtoda}. 
Moreover, a formal identification can be made between the dispersion law of 
the 1--soliton solution of the classical system and a certain branch of 
the excitation system of the quantum model, which is obtained by the Bethe 
{\it ansatz} \cite{bethetoda}. Both dispersions are 
identical in form, and in this sense the quantum analogue of a classical
soliton may be viewed as a certain stationary state of the quantum system;
see also \cite{GPM:92} for a discussion of that issue in a semiclassical 
context.
Nevertheless, such an eigenstate obtained from the Bethe {\it ansatz}
is not a dynamical object and naturally
translationally symmetric, i.~e not localized like a classical soliton. 
Moreover,
such an explicite identification is in general only possible if the
quantum and the classical system are both integrable.
Therefore the question arises whether quantum states exist which have
the essential properties of classical solitary waves, which are required
in many classical descriptions of phenomena like energy transport etc.
As such a quantum state is not translationally symmetric, it cannot be
expected to be an eigenstate of the quantum system. Moreover, its
time evolution is in general not fully coherent, but decoherence effects
due to quantum fluctuations cause a finite lifetime of such a localized
state. In the following we give an estimate for this lifetime of
semiclassical solitary waves build up from coherent states in the
quantum Toda chain.
Let us first consider the variational ground state of the Toda chain with
$\xi_{n}=\pi_{n}=0$ for all $n$. Here we clearly have $\Omega_{1}=0$,
and for $\Omega_{2}$ we find
\begin{equation}
\Omega_{2}=\frac{N}{4}(\hbar\omega)^{2}
\left(1-e^{-\gamma\left(\lambda-\frac{\gamma}{2}\Delta_{0}\right)}\right)^{2}
\,,
\end{equation} 
which is also zero for $\lambda=(\gamma/2)\Delta_{0}$. With respect to
the parameter $\eta$ entering the potential (\ref{pot}) this means 
\begin{equation}
m\omega^{2}\exp\left(-\frac{\gamma^{2}}{2}\frac{\hbar}{m\omega}\frac{1}{N}
\frac{\sin(\pi/N)}{1-\cos(\pi/N)}\right)=\eta\,.
\end{equation}
This relation determines the frequency $\omega$ which enters the
variational {\it ansatz} (\ref{ansatz}) via the phonon dispersion
$\omega(k)$. One obviously has always a non--negative solution $\omega$
for any non--negative $\eta$. Note that with this choice for $\omega$ 
the quantum corrections in the exponential factor in the variational expression
(\ref{ev2}) and also in the equations of motion cancel
with the parameter $\lambda$, but are of course present compared with the
original Hamiltonian. However, the higher order terms $\Omega_{\mu}$
with $\mu>2$ are in general non--zero for this classical ground state solution.
Thus, our variational ground state approximates the exact ground state 
within the first two orders of $\hbar$. To account for higher corrections
one has to implement a more complicated state than (\ref{ansatz}).
Therefore, in the spirit of the WKB approximation scheme we can be confident
to give a valid description of the quantum system within the first two
orders of $\hbar$.\\
Let us now turn to solitary solution to the variational equations of motion
(which are practically the same as the classical equations).
As mentioned above, such solutions do not correspond  to (approximate) 
eigenstates 
of the system like the variational ground state, but suffer decoherence 
effects in their time evolution. Nonlinear excitations in the classical Toda 
chain with periodic boundary conditions are so--called cnoidal waves which 
can be expressed in terms of Jacobi elliptic functions. As a limiting case, a 
pulse soliton arises which is given by elementary expressions \cite{Tod:81},
\begin{equation}
\pi_{n}=\pm\frac{\nu m}{\gamma}
\left(\tanh\left(\kappa(n-1)\pm\nu t\right)
-\tanh\left(\kappa n\pm\nu t\right)\right)\,,
\end{equation}
\begin{equation}
e^{-\gamma\left(\xi_{n+1}-\xi_{n}\right)}-1
=\frac{\sinh^{2}\kappa}{\cosh^{2}\left(\kappa n\pm\nu t\right)}
\end{equation}
with the soliton parameter $\kappa$, which is the inverse soliton width,
and $\nu=\omega\sinh\kappa$.
Although this solution of the variational equations of motion is, strictly
speaking, not compatible with periodic boundary conditions, it is an  
excellent numerical approximation for the cnoidal waves for large 
wave length and system size. For simplicity, we shall concentrate on the
above expressions in the following. With this solution the quantities
$\Omega_{\mu}$ can be written as
\begin{equation}
\Omega_{\mu}=(\hbar\omega)^{2}
\left(\frac{m\omega^{2}/\gamma^{2}}{\hbar\omega}\right)^{2-\mu}
Q_{\mu}(\kappa)\,,
\end{equation}
where the $Q_{\mu}$ depend only on $\kappa$. 
In particular, the $Q_{\mu}$ (and therefore the $\Omega_{\mu}$)
are time--independent since our soliton solution describes a
stationary movement, where a translation in time is equivalent to a 
translation in space. Therefore the time dependence drops out when the 
summations
over the system in the equations (\ref{O1})--(\ref{O3}) are performed.
The dimensionless quantity
$(m\omega^{2}/\gamma^{2})/(\hbar\omega)$ is the ratio of the energy
scales of the nonlinear interaction and of the linear phonon excitations.
In a semiclassical regime this quotient is large and suppresses all
orders $\Omega_{\mu}$ with $\mu>2$ (which are not considered here further,
cf. above). For $\mu=2$ we have
for an infinite system
\begin{equation}
Q_{2}(\kappa)=\frac{4\sinh^{4}\kappa}{\pi^{2}}
\sum_{l=-\infty}^{\infty}\frac{1}{\left(4l^{2}-1\right)^{2}}
\left[\sum_{n=-\infty}^{\infty}
\frac{1}{\cosh^{2}(\kappa n)\cosh^{2}(\kappa (n-l))}\right]\,.
\label{Q2}
\end{equation}
The above summations are non--elementary. The largest contribution stems
from the summand with $l=0$. Replacing the remaining sum over $n$ by an
integral, one concludes that this quantity should scale approximately
like $1/\kappa$. Indeed, a numerical evaluation of the full double sum for
$\kappa\in]0,0.5]$ shows that a very accurate value for this expression is
$(4/3\kappa)$; deviations from this occur only for large $\kappa$ and are of 
order $10^{-5}$. Therefore, we may write in a very good approximation
\begin{equation}
\Omega_{2}
=(\hbar\omega)^{2}\frac{4\sinh^{4}\kappa}{\pi^{2}}\frac{4}{3\kappa}\,,
\end{equation}
and the estimate of the decoherence time according to (\ref{ur}) is
\begin{equation}
\Delta t\geq\frac{1}{\omega}\frac{\pi\sqrt{3}}{8}
\frac{\sqrt{\kappa}}{\sinh^{2}\kappa}\,.
\end{equation}
Multiplying with the soliton velocity $c=\nu/\kappa$ one finds for the
decoherence length $\Delta l=c\Delta t$ for small $\kappa$
\begin{equation}
\Delta l\geq\frac{\pi\sqrt{3}}{8}\kappa^{-3/2}\,.
\label{cohlen}
\end{equation}
Remarkably, no system parameter or Planck's constant itself, but only
the soliton width enters  (\ref{cohlen}).
The decoherence length is large for small $\kappa$, i.~e. broad solitons.
For instance, a soliton with a width of 100 lattice units may travel 
(at least) about
ten times this distance until decoherence effects become significant.
With respect to the classical picture of solitons, this appears rather
restrictive. On the other hand, the relation (\ref{cohlen}) provides only a 
lower bound for the coherence length; e.~g. in the classical limit
$\hbar\to 0$ all decoherence effects vanish and the decoherence length 
becomes infinite.
However, for not too large values of the ratio 
$(m\omega^{2}/\gamma^{2})/(\hbar\omega)$ the decoherence length should be
assumed to be of the order of the right hand side of (\ref{cohlen}),
at least as a `conservative' estimate.


\section{Conclusions}

In this work we have examined coherent states as a time--dependent
variational {\it ansatz} for a semiclassical description of anharmonic
oscillators. In particular, the square variance of the Hamiltonian
$\langle{\cal H}^{2}\rangle-\langle{\cal H}\rangle^{2}$
within coherent states is considered. For a single anharmonic oscillator,
the first order in $\hbar$ of 
this quantity turns out to be purely given by the variational time dependence
of the quantum state, cf. equations (\ref{sv1})--(\ref{Omega1}).
Therefore, this contribution has a natural interpretation, which can
be confirmed rigorously in the case of the harmonic oscillator and 
the free particle.  
Compared with recent results on spin--coherent states \cite{ScMe:98}
this appears to be a general property of coherent states with respect
to generic quantum systems. The remaining contributions to 
$\langle{\cal H}^{2}\rangle-\langle{\cal H}\rangle^{2}$ can be used
to estimate decoherence effects which arise from quantum fluctuations.
In the foregoing section we have illustrated this by the example of the 
Toda chain. We have chosen this system, because it provides comparatively
simple expressions for the quantities considered here,
and explicite solitary solutions of the classical equations are 
available. In fact, we expect our approach to be useful for much more
general anharmonic lattices.  


\end{document}